\documentclass[amsmath,amssymb,floatfix,aps,twocolumn,superscriptaddress,nofootinbib,notitlepage,prl]{revtex4-2}

\usepackage{amsmath,amsthm,amssymb,bm,graphicx,xcolor,mathtools,enumitem}
\usepackage[colorlinks=true,urlcolor=blue,linkcolor=blue,citecolor=green,bookmarks=false]{hyperref}

\allowdisplaybreaks

\newcommand{\be}{\begin{equation}}
\newcommand{\ee}{\end{equation}}

\newcommand{\subc}{\mathrm{C}}    %sub/superscript for charge quantities
\newcommand{\subs}{\mathrm{S^z}}  %sub/superscript for spin quantities

\begin{document}

\title{Nonequilibrium full counting statistics of multicomponent strongly-interacting quantum gases with defects}

\begin{abstract}
			
We investigate the full counting statistics of a one-dimensional multicomponent impenetrable gas released from a bipartite state in the presence 
of a local defect. Spin--charge separation reduces the multivariate generating function to an exact Fredholm determinant whose kernel is built 
from time-evolved single-particle orbitals. Expressing these orbitals in terms of scattering data gives access to both transient dynamics and 
the nonequilibrium steady state. We derive the leading long-time asymptotics and identify persistent oscillations in densities and currents 
caused by multiple bound states at the junction.
We argue that interactions between components lead to a linear low-temperature correction to the current, in contrast to the quadratic correction 
in the single-component case. At equilibrium, we obtain the M-Wright distribution of spin transfer with $t^{1/4}$ scaling.

\end{abstract}

\author{Ovidiu I. P\^{a}\c{t}u} %\orcid{0000-0002-0465-8796} }
%\thanks{Both authors contributed equally to this work.}
\affiliation{Institute for Space Sciences, Bucharest-M\u{a}gurele, R 077125, Romania}

\author{Oleksandr Gamayun} %\orcid{0000-0002-2889-8487}
%\thanks{Both authors contributed equally to this work.}
\affiliation{London Institute for Mathematical Sciences, Royal Institution, 21 Albemarle, London W1S 4BS}

\author{C\u at\u alin Pa\c scu Moca}
\affiliation{Department of Physics, University of Oradea, 410087 Oradea, Romania}
\affiliation{Department of Theoretical Physics, Institute of Physics, Budapest University of Technology and Economics, H-1111 Budapest, Hungary}
\affiliation{MTA-BME Lend\"ulet ``Momentum'' Open Quantum Systems Research Group, Budapest University of Technology and Economics, H-1111 Budapest, Hungary}
	
\maketitle

%%%%%%%%%%%%%%%%%%%%%%%%%%%%%%%%%%%%%%%%%%%%%%%%%%%%%%%%%%%%%%%%%%%%%%%%%%%%%%%%%%%%%%%%%%%%%%%%%%%%%%%%%%%%%%%%%%%%%%%%%%%%%%%%%%%%%%%%%%%%%%%%
%%%%%%%%%%%%%%%%%%%%%%%%%%%%%%%%%%%%%%%%%%%%%%%%%%%%%%%%%%%%%%%%%%%%%%%%%%%%%%%%%%%%%%%%%%%%%%%%%%%%%%%%%%%%%%%%%%%%%%%%%%%%%%%%%%%%%%%%%%%%%%%%
%%%%%%%%%%%%%%%%%%%%%%%%%%%%%%%%%%%%%%%%%%%%%%%%%%%%%%%%%%%%%%%%%%%%%%%%%%%%%%%%%%%%%%%%%%%%%%%%%%%%%%%%%%%%%%%%%%%%%%%%%%%%%%%%%%%%%%%%%%%%%%%%

\paragraph{Introduction.---}

Full counting statistics (FCS) \cite{MHM09}, defined through the generating function of the cumulants of a random variable, encodes the  full 
probability distribution of an observable, rather than only its expectation value. Access to multipoint correlations and fluctuations provided by 
the FCS, together with its experimental realization in cold atomic gases \cite{ETSA06,AJKB10,JABK11,SKEM17,WRYM22} and on quantum computing 
platforms \cite{YD24,Google24,SMKV24}, has renewed interest in its computation in one-dimensional quantum many-body systems. Analytical results 
for FCS have been obtained for free quadratic systems \cite{LL93,ER13,DLSB14,Yosh18,DCVA20,GLC20,DBVA21,GZI23}, conformal field theories \cite{BD16}, 
and interacting classical and quantum systems using a variety of techniques \cite{LF08,CEG17,SP17,GEC18,AG19,BPC18,PPG19,SDMS21,ARV21,MNGV23,KSSP24,
GMVK24,CR24,KSPI22,AMWK25,NKT26}. Furthermore, FCS has been shown to distinguish quantum phases \cite{IA13,WZZZ24} and to characterize dynamical 
universality classes \cite{KSSP24,GMVK24}.

% In this manuscript, we derive an exact multivariate FCS for a one-dimensional multicomponent gas of impenetrable particles in the continuum. The 
% two halves of the system are initially thermalized independently and separated by a hard wall. At $t=0$, we remove the wall and quench the 
% localized potential from $V_I(z)$ to $V_F(z)$, thereby joining the two subsystems. This partitioning protocol probes transport through a general 
% defect while allowing different temperatures and component-dependent chemical potentials on the two sides.

We address this question for a one-dimensional $\kappa$-component impenetrable gas prepared in a bipartite quench. Two initially thermalized halves 
are separated by a hard wall; at $t=0$ the wall is removed and the localized potential is quenched from $V_I(z)$ to $V_F(z)$. The protocol provides 
a controlled setting for defect-mediated transport between reservoirs with different temperatures and chemical potentials.

Our main result is an exact Fredholm determinant for the component-resolved FCS. To compute it, we use spin--charge separation; at infinite repulsion, 
the charge-sector wave function evolves as that of  spinless free fermions, whereas the dynamics in the spin sector is  frozen. We use this property 
to perform the grand-canonical sum and  obtain an 
exact Fredholm determinant for the component-resolved FCS. Its kernel is expressed in terms of single-particle orbitals constructed from the 
Jost solutions and scattering data of $V_I$ and $V_F$.
This framework enables both a full numerical analysis of the transient dynamics and the analytical derivation of particle densities and currents 
in the nonequilibrium steady state. We show that, in the absence of bound states of $V_F$, the leading large-time behavior of the FCS is governed 
by a Levitov--Lesovik-like formula (LL) \cite{LL93,Scho07}. In contrast, when multiple bound states are supported, both densities and particle 
currents exhibit oscillatory behavior near the junction, with frequencies set by the corresponding energy differences \cite{GLC20,GZI23}.
We further uncover anomalous low-temperature behavior specific to strongly interacting multicomponent systems. Whereas steady-state observables in 
single-component systems receive the familiar quadratic temperature corrections, balanced multicomponent systems exhibit linear temperature 
corrections. Finally, we relate the LL asymptotics to anomalous spin transport 	\cite{KSPI22,KVZ22,GMVK24,GMV24,KSSP24,KSIP24,KIPH25,YK25,YK26,
Fujimoto2026,YKBI26,UVGN26,POZS26} in equilibrium.

%%%%%%%%%%%%%%%%%%%%%%%%%%%%%%%%%%%%%%%%%%%%%%%%%%%%%%%%%%%%%%%%%%%%%%%%%%%%%%%%%%%%%%%%%%%%%%%%%%%%%%%%%%%%%%%%%%%%%%%%%%%%%%%%%%%%%%%%%%%%%%%%
%%%%%%%%%%%%%%%%%%%%%%%%%%%%%%%%%%%%%%%%%%%%%%%%%%%%%%%%%%%%%%%%%%%%%%%%%%%%%%%%%%%%%%%%%%%%%%%%%%%%%%%%%%%%%%%%%%%%%%%%%%%%%%%%%%%%%%%%%%%%%%%%
%%%%%%%%%%%%%%%%%%%%%%%%%%%%%%%%%%%%%%%%%%%%%%%%%%%%%%%%%%%%%%%%%%%%%%%%%%%%%%%%%%%%%%%%%%%%%%%%%%%%%%%%%%%%%%%%%%%%%%%%%%%%%%%%%%%%%%%%%%%%%%%%	

\paragraph{Model and setup.---}

We consider a $\kappa$-component system of impenetrable one-dimensional particles in the presence of a local, time-dependent potential. In 
the second-quantization formalism, the Hamiltonian reads
\begin{align*}
H = & \int dz\, \frac{\hbar^2}{2m}
\left( \partial_z \boldsymbol{\Psi}^\dagger \, \partial_z \boldsymbol{\Psi} \right)
+ \frac{\mathfrak{g}}{2} :\!\left(\boldsymbol{\Psi}^\dagger \boldsymbol{\Psi}\right)^2\!:
+ V(z,t)\, \boldsymbol{\Psi}^\dagger \boldsymbol{\Psi}\, .
\end{align*}
Here, $m$ denotes the particle mass, and $\mathfrak{g}=\infty$ denotes the strength of the repulsive interaction. The field operators are defined 
as $\boldsymbol{\Psi}(z)=\left(\Psi_1(z),\ldots,\Psi_\kappa(z)\right)^T$ and 
$\boldsymbol{\Psi}^\dagger(z)=\left(\Psi_1^\dagger(z),\ldots,\Psi_\kappa^\dagger(z)\right)$,
where $\Psi_\sigma(z)$, $\sigma\in\{1,\ldots,\kappa\}$, are fermionic or bosonic field operators. These operators satisfy the (anti)commutation 
relations
$
\Psi_\sigma(z)\Psi_{\sigma'}^\dagger(z') - \epsilon\, \Psi_{\sigma'}^\dagger(z')\Psi_\sigma(z)
= 
\delta_{\sigma\sigma'}\,\delta(z-z'),
$
with $\epsilon=-1$ for fermions and $\epsilon=+1$ for bosons.

We consider a general nonequilibrium bipartite setup in which a hard wall is imposed at $z=0$, in addition to the local external potential. The 
system is thus split into two independent subsystems ($z<0$ and $z>0$), which are prepared for $t<0$ in states with possibly different particle 
numbers, temperatures, or chemical potentials. At $t=0$, the barrier is removed and the external potential is quenched. Specifically, the 
potential takes the form $V(z,t)=V_I(z)$ (the initial hard wall at $z=0$ is imposed in addition to $V_I(z)$) for $t\le 0$ and $V(z,t)=V_F(z)$ 
for $t>0$. The initial potential $V_I(z)$ is assumed to vanish sufficiently rapidly as $|z|\to\infty$ and not to support bound states. The final 
potential $V_F(z)$ satisfies $\int_{-\infty}^{\infty} (1+|z|)\,|V_F(z)|\,dz < \infty, $ which ensures that the number of bound states is finite. 
We restrict to potentials vanishing outside a finite interval. We denote the initial Hamiltonian by $H_I = H(t\le 0)$ and the final Hamiltonian, 
which dictates the post-quench time evolution, by $H_F = H(t>0)$.

%%%%%%%%%%%%%%%%%%%%%%%%%%%%%%%%%%%%%%%%%%%%%%%%%%%%%%%%%%%%%%%%%%%%%%%%%%%%%%%%%%%%%%%%%%%%%%%%%%%%%%%%%%%%%%%%%%%%%%%%%%%%%%%%%%%%%%%%%%%%%%%%
%%%%%%%%%%%%%%%%%%%%%%%%%%%%%%%%%%%%%%%%%%%%%%%%%%%%%%%%%%%%%%%%%%%%%%%%%%%%%%%%%%%%%%%%%%%%%%%%%%%%%%%%%%%%%%%%%%%%%%%%%%%%%%%%%%%%%%%%%%%%%%%%
%%%%%%%%%%%%%%%%%%%%%%%%%%%%%%%%%%%%%%%%%%%%%%%%%%%%%%%%%%%%%%%%%%%%%%%%%%%%%%%%%%%%%%%%%%%%%%%%%%%%%%%%%%%%%%%%%%%%%%%%%%%%%%%%%%%%%%%%%%%%%%%%

\paragraph{Full counting statistics.---}

Our main object of interest is the multivariate full counting statistics (FCS) of the particle numbers in an arbitrary interval $[x,y]$. Let
$ N_{[x,y],\sigma}(t)= e^{i t H_F} \int_x^y n_\sigma(z) dz\, e^{-i t H_F}$ denote the number of particles of component $\sigma$ in the interval 
$[x,y]$, where $n_\sigma(z)=\Psi_\sigma^\dagger(z)\Psi_\sigma(z)$. We use the moment-generating function
\begin{equation}\label{fcs}
\chi_{[x,y]}(\boldsymbol{\lambda}; t)
=
\left \langle e^{\sum_{\sigma=1}^{\kappa} \lambda_\sigma N_{[x,y],\sigma}(t)} \right\rangle \, .
\end{equation}
The expectation value is taken with respect to the grand-canonical density matrix of the initial Hamiltonian $H_I$, while the system 
subsequently  evolves under $H_F$. The characteristic function follows by setting $\boldsymbol{\lambda}\to i\boldsymbol{\lambda}$.
	
The characteristic function  encodes the full probability distribution of the particle numbers of each component in the interval 
$[x,y]$  and therefore completely determines the corresponding transport properties. In particular, the moments are obtained (with the interval 
subscript suppressed whenever there is no risk of confusion) as
$
\left\langle \big(N_\sigma(t)\big)^m \right\rangle
=
\left.\frac{\partial^m}{\partial \lambda_\sigma^m}\chi(\boldsymbol{\lambda}; t)\right|_{\boldsymbol{\lambda}=0}.
$
More generally, (\ref{fcs}) generates density correlations, from which multipoint correlations, particle currents, and emptiness 
formation probabilities (EFPs)~\cite{KBI93} can be extracted. For instance, the time-dependent density and density--density correlations are 
given by
$
\langle n_\sigma(y,t) \rangle
=
\left. \frac{\partial}{\partial y} \frac{\partial}{\partial \lambda_\sigma} \chi(\boldsymbol{\lambda}; t) \right|_{\boldsymbol{\lambda}=0},
$
$
\langle n_\sigma(y,t) n_\sigma(x,t) \rangle
= 
\left. -\frac{1}{2} \frac{\partial^2}{\partial x \partial y} \frac{\partial^2}{\partial \lambda_\sigma^2} \chi(\boldsymbol{\lambda}; t)
\right|_{\boldsymbol{\lambda}=0},
$
with $\sigma\in\{1,\ldots,\kappa\}$. The continuity equation $\partial_t n_\sigma(z,t) + \partial_z j_\sigma(z,t) = 0$ can then be used to obtain 
correlation functions of the particle current. % $j_\sigma(x,t)$. 

%%%%%%%%%%%%%%%%%%%%%%%%%%%%%%%%%%%%%%%%%%%%%%%%%%%%%%%%%%%%%%%%%%%%%%%%%%%%%%%%%%%%%%%%%%%%%%%%%%%%%%%%%%%%%%%%%%%%%%%%%%%%%%%%%%%%%%%%%%%%%%%%
%%%%%%%%%%%%%%%%%%%%%%%%%%%%%%%%%%%%%%%%%%%%%%%%%%%%%%%%%%%%%%%%%%%%%%%%%%%%%%%%%%%%%%%%%%%%%%%%%%%%%%%%%%%%%%%%%%%%%%%%%%%%%%%%%%%%%%%%%%%%%%%%
%%%%%%%%%%%%%%%%%%%%%%%%%%%%%%%%%%%%%%%%%%%%%%%%%%%%%%%%%%%%%%%%%%%%%%%%%%%%%%%%%%%%%%%%%%%%%%%%%%%%%%%%%%%%%%%%%%%%%%%%%%%%%%%%%%%%%%%%%%%%%%%%	

\paragraph{Determinant representation of the FCS.---}
	
Computing the FCS in nonequilibrium settings is a notoriously challenging task, even for single-component systems, and becomes significantly 
more involved in multicomponent systems. The characteristic function~\eqref{fcs} can be evaluated by employing spin--charge separation in the 
impenetrable limit and using the fact that the dynamics is entirely governed by the charge sector.  Although the expectation value remains 
highly nontrivial, the grand-canonical sum over all many-body states can be performed exactly in several cases using von Koch's determinant 
formula~\cite{Born10}, as detailed below. We first introduce some notation.
	
For a finite system, we impose hard-wall boundary conditions. More precisely, let us denote the single-particle eigenfunctions (orbitals) of the initial 
charge Hamiltonian $H_I^{\mathrm{SP}} = -\frac{\hbar^2}{2m}\frac{\partial^2}{\partial z^2}+V_I(z)$ by $\Lambda_{S,\epsilon}$, where the index 
$S\in\{L,R\}$ labels the left or right part of the system, i.e., $H_I^{\mathrm{SP}}\Lambda_{S,\epsilon} = \epsilon \Lambda_{S,\epsilon}$. 
The hard-wall conditions at the junction and the edges of the system produce a discrete set of energies $\epsilon$. The post-quench time 
dependence of the orbitals is governed by the single-particle Hamiltonian $ H_F^{\mathrm{SP}} = -\frac{\hbar^2}{2m}
\frac{\partial^2}{\partial z^2}+V_F(z) $, i.e., $\Lambda_{S,\epsilon}(z,t) = e^{it H_F^{\mathrm{SP}}}\Lambda_{S,\epsilon}(z, t=0)$. This 
convention evolves the complex conjugate of the usual Schr\"odinger orbital.
	
The subsystems are each initially prepared in thermal equilibrium at inverse temperatures $\beta_L$ and $\beta_R$, and with chemical potentials 
$ \boldsymbol{\mu}_L=(\mu_{L,1},\ldots,\mu_{L,\kappa})$, $ \boldsymbol{\mu}_R=(\mu_{R,1},\ldots,\mu_{R,\kappa})$, respectively. 
For each energy, we introduce the generalized Fermi occupation factors
\begin{equation}\label{fermifactor}
\theta_S(\epsilon)=\sum_{\sigma=1}^{\kappa} \theta^\sigma_S(\epsilon),\quad 
\theta^\sigma_S(\epsilon) =
\frac{e^{-\beta_S \epsilon} e^{\beta_S \mu_{S,\sigma}} }{1 + \sum_{\sigma} e^{\beta_S (\mu_{S,\sigma}-\epsilon)}} \, ,
\end{equation}
and define a coefficient depending on the counting fields as
\begin{equation}\label{defcoeff}
\xi_S(\boldsymbol{\lambda})
=
\frac{\sum_{\sigma=1}^{\kappa} e^{\lambda_\sigma} e^{\beta_S \mu_{S,\sigma}}}{\sum_{\sigma=1}^{\kappa} e^{\beta_S \mu_{S,\sigma}}}.
\end{equation}

We can now state our result for the FCS in the following cases:
	
(A) One subsystem is characterized by arbitrary $\beta_S$ and $\boldsymbol{\mu}_S$, while the other subsystem is empty, $\theta_{S'}(\epsilon)=0$ with 
$S'\neq S$.
	
(B) Both subsystems are characterized by arbitrary temperatures $\beta_L$ and $\beta_R$ but balanced chemical potentials,  
$\boldsymbol{\mu}_L=(\mu_L,\ldots,\mu_L)$ and $\boldsymbol{\mu}_R=(\mu_R,\ldots,\mu_R)$. 
	
(C) Both subsystems are characterized by arbitrary temperatures and chemical potentials, but the object of interest is the total number of 
particles irrespective of spin, i.e., $\boldsymbol{\lambda}=(\lambda,\ldots,\lambda)$.
	
In each of these cases, the FCS \eqref{fcs} admits the following Fredholm determinant representation
\begin{align}\label{Lenard}
\chi_{[x,y]}(\boldsymbol{\lambda}; t)=
\det\!\left[ \boldsymbol{1} +P_{[x,y]} \sum_{S=L,R} \bigl(\xi_S(\boldsymbol{\lambda})-1\bigr) \hat{K}_S \right],
\end{align}
where $\hat{K}_S$ are integral operators acting on the real line with kernels
\begin{align}\label{kerLenard}
K_S(z,z';t)=\sum_{\epsilon=1}^{\infty} \theta_S(\epsilon)\, \Lambda_{S,\epsilon}^*(z,t)\, \Lambda_{S,\epsilon}(z',t),
\end{align}
and $P_{[x,y]}$ is the projector on the interval $[x,y]$. 
In the thermodynamic limit, the sum over energies is replaced by a discrete sum over bound states and an integral over the continuous spectrum, 
$\sum_\epsilon \to 2\int_0^\infty dq/\pi$, with $\epsilon \to \epsilon_q = q^2/\gamma$, where $\gamma=2m/\hbar^2$. Henceforth, we 
set $\hbar=k_B=1$.
	
The orbitals can be expressed in terms of scattering data. These data are analytically available for a broad class of potentials and can 
otherwise be computed efficiently numerically, as demonstrated for a single-component system with particles initially present in only one 
subsystem \cite{GZI23}. The key ingredients are the Jost solutions \cite{NMPZ84} of the spinless scattering problem. The Jost functions 
corresponding to the final potential $V_F(z)$ are denoted by $\psi_k(z)$ and $\varphi_k(z)$, and those corresponding to the initial potential 
$V_I(z)$ by $\Psi_k(z)$ and $\Phi_k(z)$. The Jost solutions are defined by their asymptotic behavior 
\begin{align}\label{as1}
\psi_k(z), \Psi_k(z) & = e^{-i k z}+o(1), \quad z\to +\infty,   \\ 
\varphi_k(z), \Phi_k(z)& =e^{-i k z}+o(1), \quad z\to -\infty. \label{as2}
\end{align}
The relation $\varphi_k=a_k\psi_k+b_k\psi_{-k}$ defines the scattering coefficients, with transmission $T(k)=1/|a_k|^2$ and reflection 
$R(k)=|b_k|^2/|a_k|^2$.  The zeros $a_{i\varkappa_n}=0$, $\varkappa_n>0$, in the upper half-plane correspond to bound-state energies 
$E_n^b=-\varkappa_n^2/\gamma$.
	
At $t=0$ the single-particle orbitals can be written as
\begin{align}\label{defL}
\Lambda_{L,\epsilon_q}(z) & = \Phi_q(0){\rm Im}\left[\Phi_q(z)/\Phi_q(0)\right],     \\
\Lambda_{R,\epsilon_q}(z) & = \Psi^*_q(0){\rm Im}\left[\Psi_q(z)/\Psi_q(0)\right].
\end{align}
Then the large-time asymptotic expansion of the evolved orbitals takes the form
\begin{align}\label{evLambda}
\Lambda_{L,\epsilon_q}(z,t) & = \frac{\psi_q(z)e^{it\epsilon_q}}{2i a^*_q} + \sum_{n=1}^{N_b} B^{(n)}_{L,q} \psi^*_{i\varkappa_n}(z)e^{i t E^b_n},     \\
\Lambda_{R,\epsilon_q}(z,t) & =  \frac{i\varphi^*_q(z)e^{it\epsilon_q}}{2a_q}+ \sum_{n=1}^{N_b} B^{(n)}_{R,q} \varphi_{i\varkappa_n}(z)e^{i t E^b_n},
\end{align}
up to corrections of order $\mathcal{O}(t^{-1/2})$. The coefficients $B^{(n)}_{S,q}$ can be expressed through integrals of the pre- and 
post-quench Jost functions and the difference between the initial and final potentials. Their exact forms are given in Sec.~\ref{timeevol} of 
the End Matter.

%%%%%%%%%%%%%%%%%%%%%%%%%%%%%%%%%%%%%%%%%%%%%%%%%%%%%%%%%%%%%%%%%%%%%%%%%%%%%%%%%%%%%%%%%%%%%%%%%%%%%%%%%%%%%%%%%%%%%%%%%%%%%%%%%%%%%%%%%%%%%%%%
%%%%%%%%%%%%%%%%%%%%%%%%%%%%%%%%%%%%%%%%%%%%%%%%%%%%%%%%%%%%%%%%%%%%%%%%%%%%%%%%%%%%%%%%%%%%%%%%%%%%%%%%%%%%%%%%%%%%%%%%%%%%%%%%%%%%%%%%%%%%%%%%
%%%%%%%%%%%%%%%%%%%%%%%%%%%%%%%%%%%%%%%%%%%%%%%%%%%%%%%%%%%%%%%%%%%%%%%%%%%%%%%%%%%%%%%%%%%%%%%%%%%%%%%%%%%%%%%%%%%%%%%%%%%%%%%%%%%%%%%%%%%%%%%%	

\begin{figure}
\centering
\includegraphics[width=1\linewidth]{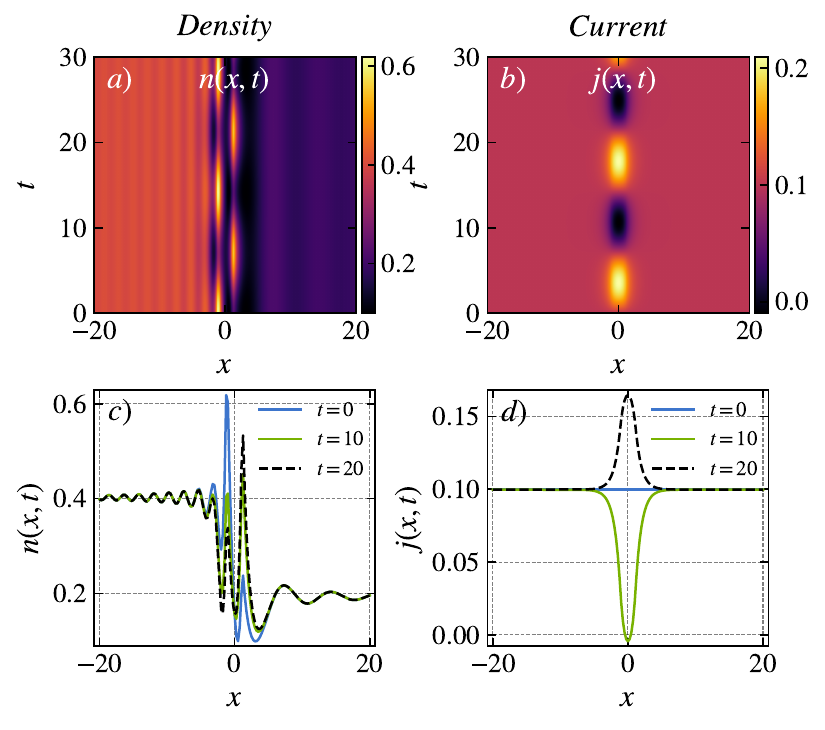}
\caption{
Total density (a) and particle current (b) computed from the asymptotic expressions (\ref{dens}) and (\ref{current}) for a two-component system in 
the presence of a double $\delta$ barrier with $g=-1.3$ and $d=2.3$. The system is initially prepared in a bipartite setup with 
$(\beta_L,\beta_R)=(100,100)$, $\boldsymbol{\mu}_L=(2,2)$, and $\boldsymbol{\mu}_R=(0.2,0.2)$. Panels (c) and (d) show spatial profiles at times 
$t=0,10,20$.
}\label{fig1}
\end{figure}

%%%%%%%%%%%%%%%%%%%%%%%%%%%%%%%%%%%%%%%%%%%%%%%%%%%%%%%%%%%%%%%%%%%%%%%%%%%%%%%%%%%%%%%%%%%%%%%%%%%%%%%%%%%%%%%%%%%%%%%%%%%%%%%%%%%%%%%%%%%%%%%%
%%%%%%%%%%%%%%%%%%%%%%%%%%%%%%%%%%%%%%%%%%%%%%%%%%%%%%%%%%%%%%%%%%%%%%%%%%%%%%%%%%%%%%%%%%%%%%%%%%%%%%%%%%%%%%%%%%%%%%%%%%%%%%%%%%%%%%%%%%%%%%%%
%%%%%%%%%%%%%%%%%%%%%%%%%%%%%%%%%%%%%%%%%%%%%%%%%%%%%%%%%%%%%%%%%%%%%%%%%%%%%%%%%%%%%%%%%%%%%%%%%%%%%%%%%%%%%%%%%%%%%%%%%%%%%%%%%%%%%%%%%%%%%%%%	

\paragraph{Densities and particle currents.---}

In the long-time limit following the quench, the system approaches an asymptotic regime consisting of a nonequilibrium steady-state (NESS) 
contribution supplemented by localized oscillatory terms when multiple bound states are present.
Combining the representation~\eqref{Lenard} with the orbital asymptotics~\eqref{evLambda}, we obtain %, for $\sigma=1,\ldots,\kappa$,%
\begin{align}\label{dens}
\langle n_\sigma(x,t)\rangle
&=
n_\sigma(x) + \sum_{n=1}^{N_b} n_{\sigma,n}^{B}(x) \nonumber\\
&\qquad + \sum_{m<n}^{N_b} n_{\sigma,m,n}^{BB}(x) \cos\!\left[ t(E^b_m-E^b_n)  \right],
\end{align}
and
\begin{align}\label{current}
\langle j_\sigma(x,t) \rangle
=
j_\sigma + \sum_{m<n}^{N_b} J_{\sigma,m,n}^{BB}(x) \sin\!\left[ t(E^b_m-E^b_n) \right].
\end{align}
The steady-state density contribution is given by
\begin{align}\label{dens2}
n_\sigma(x)
=
\int_0^\infty \frac{dq}{2\pi} \left[\theta^\sigma_L(\epsilon_q)\,|\psi_q(x)|^2 +\theta^\sigma_R(\epsilon_q)\,|\varphi_q(x)|^2 \right] T(q),
\end{align}
while the steady current is given by the Landauer--B\"uttiker formula \cite{Land57,Land70,Butt86},
\begin{align}\label{jLB}
j_\sigma
=
\frac{1}{\gamma} \int_0^\infty \frac{dq}{\pi} \bigl[ \theta_L^\sigma(\epsilon_q) - \theta_R^\sigma(\epsilon_q) \bigr] q\,T(q).
\end{align}
Thus, the steady-state contributions depend solely on the Jost solutions and the transmission coefficient $T(q)=1/|a_q|^{2}$ of the post-quench 
potential. The position dependence of the coefficients $n_{\sigma,n}^{B}(x)$, $n_{\sigma,m,n}^{BB}(x)$, and $J_{\sigma,m,n}^{BB}(x)$ is 
determined solely by the bound-state wave functions; these coefficients are therefore localized near the junction. Their explicit forms can be 
obtained once $B^{(n)}_{S,q}$ are known (see Sec.~\ref{curandden} of the End Matter).
Exact analytical expressions are available for a broad class of local potentials with known scattering data, including square wells, 
$\delta$-barriers, the P\"oschl--Teller \cite{PT33}, Eckart \cite{Eckar30}, Morse \cite{Mors29}, and Rosen--Morse potentials \cite{RM32}. 
In particular, for a single $\delta$ barrier in the single-component limit, we recover the results of Ref.~\cite{GDS22}. 
For the quench from $V_I=0$ to the double-well potential $ V_F(z)=g\,\delta(z+d/2)+g\,\delta(z-d/2)$, we present the results in Fig.~\ref{fig1} 
for a two-component system with parameters given in the caption. They correspond to the regime $|g d|>2$, in which two bound states are present 
in the spectrum. The persistent temporal oscillations are quite pronounced (cf. Refs.~\cite{GLC20,GZI23}). 
Another notable feature is the presence of spatial oscillations in the particle density away from the junction [see Fig.~\ref{fig1}(a) 
and Fig.~\ref{fig1}(c)], which can be interpreted as nonequilibrium analogs of Friedel oscillations~\cite{GDS22}. This effect is progressively 
suppressed as the temperature increases.
	
Equations~\eqref{dens}, \eqref{dens2}, \eqref{current}, and \eqref{jLB} reveal that bound states generically induce persistent local oscillations 
in both densities and currents, while the NESS contribution retains the universal Landauer--Büttiker form. Moreover, the oscillatory terms 
retain memory of the prequench state.

%%%%%%%%%%%%%%%%%%%%%%%%%%%%%%%%%%%%%%%%%%%%%%%%%%%%%%%%%%%%%%%%%%%%%%%%%%%%%%%%%%%%%%%%%%%%%%%%%%%%%%%%%%%%%%%%%%%%%%%%%%%%%%%%%%%%%%%%%%%%%%%%
%%%%%%%%%%%%%%%%%%%%%%%%%%%%%%%%%%%%%%%%%%%%%%%%%%%%%%%%%%%%%%%%%%%%%%%%%%%%%%%%%%%%%%%%%%%%%%%%%%%%%%%%%%%%%%%%%%%%%%%%%%%%%%%%%%%%%%%%%%%%%%%%
%%%%%%%%%%%%%%%%%%%%%%%%%%%%%%%%%%%%%%%%%%%%%%%%%%%%%%%%%%%%%%%%%%%%%%%%%%%%%%%%%%%%%%%%%%%%%%%%%%%%%%%%%%%%%%%%%%%%%%%%%%%%%%%%%%%%%%%%%%%%%%%%	

\paragraph{Low-temperature corrections.---}

Another distinctive feature of the NESS in strongly interacting multicomponent systems is the low-temperature behavior of transport observables, 
which can be traced to the form of the generalized Fermi functions~\eqref{fermifactor}. Although these distributions resemble the conventional 
Fermi function of a single-component system, their low-temperature properties are qualitatively different. This difference is already apparent 
in the Sommerfeld expansion of observables such as the particle current. For example, the low-temperature expansion of the Landauer--B\"uttiker 
current~\eqref{jLB} takes the form 
\begin{widetext}
\begin{align}\label{lowt}
j_\sigma
&=
\frac{1}{\kappa} \left(\int_0^{\mu_L}\frac{T(\varepsilon)}{2\pi}\, d\varepsilon -\int_0^{\mu_R}\frac{T(\varepsilon)}{2\pi}\, d\varepsilon\right)
+
\frac{\ln \kappa}{\kappa}\left(\frac{T(\mu_L)}{2\pi}\frac{1}{\beta_L} - \frac{T(\mu_R)}{2\pi}\frac{1}{\beta_R}\right)\nonumber\\
&\quad
+
\frac{1}{2} \left[ \frac{(\ln \kappa)^2}{\kappa} + \frac{\pi^2}{3\kappa}\right]
\left(\frac{T'(\mu_L)}{2\pi}\frac{1}{\beta_L^2} - \frac{T'(\mu_R)}{2\pi}\frac{1}{\beta_R^2} \right)
+
\mathcal{O}\!\left(\beta_L^{-3},\beta_R^{-3}\right)\, ,
\end{align}
\end{widetext}
where $T'(\varepsilon)$ represents the derivative of the transmission coefficient with respect to energy. In striking contrast to 
single-component systems, where steady-state observables receive only quadratic low-temperature corrections \cite{Pendry1983,BD16,GDS22}, 
strongly interacting multicomponent systems exhibit a linear temperature dependence. This anomalous behavior originates directly from the 
structure of the generalized Fermi functions~\eqref{fermifactor}.

%%%%%%%%%%%%%%%%%%%%%%%%%%%%%%%%%%%%%%%%%%%%%%%%%%%%%%%%%%%%%%%%%%%%%%%%%%%%%%%%%%%%%%%%%%%%%%%%%%%%%%%%%%%%%%%%%%%%%%%%%%%%%%%%%%%%%%%%%%%%%%%%
%%%%%%%%%%%%%%%%%%%%%%%%%%%%%%%%%%%%%%%%%%%%%%%%%%%%%%%%%%%%%%%%%%%%%%%%%%%%%%%%%%%%%%%%%%%%%%%%%%%%%%%%%%%%%%%%%%%%%%%%%%%%%%%%%%%%%%%%%%%%%%%%
%%%%%%%%%%%%%%%%%%%%%%%%%%%%%%%%%%%%%%%%%%%%%%%%%%%%%%%%%%%%%%%%%%%%%%%%%%%%%%%%%%%%%%%%%%%%%%%%%%%%%%%%%%%%%%%%%%%%%%%%%%%%%%%%%%%%%%%%%%%%%%%%

\paragraph{Long-time asymptotics.---}

We next consider the FCS of the charge transferred to the right subsystem. 
Formally, this corresponds to the case $[x,y]=[0,\infty)$, for which expression~\eqref{Lenard} ceases to define 
a Fredholm determinant because the kernel is no longer trace class.
It remains well defined in case (A), when the right part of the system is initially empty, or in case (C), with 
the definition in Eq.~\eqref{fcs} modified to
\begin{equation}\label{LL00}
\chi_R(\boldsymbol{\lambda}; t)
=
\left\langle
e^{\sum_{\sigma=1}^{\kappa} \lambda_\sigma N_{R,\sigma}(t)}
e^{-\sum_{\sigma=1}^{\kappa} \lambda_\sigma N_{R,\sigma}(0)}
\right\rangle. 
\end{equation}
In case (C), instead of Eq.~\eqref{Lenard}, one obtains
\begin{equation}\label{LL0}
\chi_R(\boldsymbol{\lambda}; t) =  \det\!\left[ \boldsymbol{1} +P_{R} \bigl(e^{\lambda}-1\bigr) \hat{K}_L + P_{L} \bigl(e^{-\lambda}-1\bigr) 
\hat{K}_R\right]
\end{equation}
where $P_L$ ($P_R$) is the projector onto the left (right) subsystem, corresponding in this case to $x<0$ ($x>0$).
	
For cases (A) and (C), in the absence of bound states and for smooth occupation functions $\theta_S(\epsilon_q)$, the leading 
long-time term can be obtained by solving the associated Riemann--Hilbert problem~\cite{KKMS09,Slav10,Kozl12,GKM26} or by 
employing effective form-factor techniques~\cite{GIZ21,CG22,ZNIG22,GZ25}:
\begin{align}\label{LL}
\ln \chi_R(\boldsymbol{\lambda};t)
=
\frac{t}{2\pi} \int_0^\infty dq\, \frac{2 q}{\gamma} \ln\!\left[ 1+T(q)L(\xi(\boldsymbol{\lambda}),q) \right],
\end{align}
where $ L(\xi,q) = (\xi-1)\theta_L(\epsilon_q)(1-\theta_R(\epsilon_q)) + (\xi^{-1}-1)\theta_R(\epsilon_q)(1-\theta_L(\epsilon_q)). $
Notice that in case (A), because the right subsystem is initially empty, we set $\theta_R(\epsilon_q)=0$, while in case (C), 
$\xi(\boldsymbol{\lambda})=e^\lambda$ for the FCS of the transferred charge, with $\boldsymbol{\lambda}=(\lambda,\ldots,\lambda)$.
Here, $T(q)$ denotes the transmission coefficient associated with the post-quench potential. If the occupation functions $\theta_S(\epsilon_q)$ exhibit 
discontinuities, as is the case at zero temperature, additional logarithmic corrections in time arise.
Within these conditions, Eq.~\eqref{LL} generalizes the Levitov--Lesovik formula~\cite{LL93,Scho07}, originally derived for spinless free fermions, 
to impenetrable multicomponent particles. Because the impenetrable charge sector is the same for bosons and fermions, the formula has the same 
form for both statistics; this differs from the free-particle case~\cite{Klich2003}.

For balanced equilibrium states, $\beta_L=\beta_R=\beta$ and $\boldsymbol{\mu}_L=\boldsymbol{\mu}_R=(\mu,\ldots,\mu)$, with perfect 
transmission $T(q)=1$, the charge-transfer asymptotics implies anomalous spin-current fluctuations 
\cite{KSPI22,KVZ22,GMVK24,GMV24,KSSP24,KSIP24,KIPH25,YK25,YK26,Fujimoto2026,YKBI26,UVGN26,POZS26}. In the long-time scaling limit, 
the spin-transfer distribution is the M-Wright distribution \cite{Wright1933,MMP10}
$
\mathbb{P}_{\text{MW}}[\mathcal{J},\sigma]=\int_0^\infty ds\, e^{-s^2/(2\sigma^2)-\mathcal{J}^2/(2s)}/(\pi\sigma\sqrt{s}).
$
For an arbitrary traceless Hermitian Cartan generator of $SU(\kappa)$, $\Lambda_z=\text{diag}(q_1,\ldots,q_\kappa)$, with $\sum_{\alpha=1}^\kappa 
q_\alpha=0$, the corresponding spin (magnetization) in the right subsystem is $ S_R^z(t)=\int_0^\infty\sum_{\alpha=1}^\kappa 
q_\alpha n_\alpha(x,t)\,dx. $
Denoting by $\mathbb{P}[\Delta S_R^z,t]$ the probability distribution of the integrated spin current across the junction, where 
$\Delta S_R^z=S_R^z(t)-S_R^z(0)$, one finds, using the spin-charge subordination mechanism introduced in \cite{MPZD26},  that in the long-time limit the distribution takes the scaling form ($\Delta S_R^z=\mathcal{J}_s t^{1/4}$)
\begin{equation}\label{MW1}
\lim_{t\rightarrow\infty}\mathbb{P}[t^{1/4} \mathcal{J}_s,t]=\frac{1}{t^{1/4}}\frac{1}{\sigma_{\subs}}
\mathbb{P}_{\text{MW}}\left[\frac{\mathcal{J}_s}{\sigma_{\subs}},\sigma_{\subc}\right]
\end{equation}
with
\begin{equation}\label{MW2}
\sigma_{\subc}=\left(\frac{1}{\pi\beta}\frac{\kappa}{\kappa+e^{-\beta\mu}}\right)^{1/2},
\ 
\sigma_{\subs}=\left(\sum_{\alpha=1}^\kappa \frac{q_\alpha^2}{\kappa}\right)^{1/2}.
\end{equation}
Some details are outlined in Sec.~\ref{anofluctu} of the End Matter.

%%%%%%%%%%%%%%%%%%%%%%%%%%%%%%%%%%%%%%%%%%%%%%%%%%%%%%%%%%%%%%%%%%%%%%%%%%%%%%%%%%%%%%%%%%%%%%%%%%%%%%%%%%%%%%%%%%%%%%%%%%%%%%%%%%%%%%%%%%%%%%%%
%%%%%%%%%%%%%%%%%%%%%%%%%%%%%%%%%%%%%%%%%%%%%%%%%%%%%%%%%%%%%%%%%%%%%%%%%%%%%%%%%%%%%%%%%%%%%%%%%%%%%%%%%%%%%%%%%%%%%%%%%%%%%%%%%%%%%%%%%%%%%%%%
%%%%%%%%%%%%%%%%%%%%%%%%%%%%%%%%%%%%%%%%%%%%%%%%%%%%%%%%%%%%%%%%%%%%%%%%%%%%%%%%%%%%%%%%%%%%%%%%%%%%%%%%%%%%%%%%%%%%%%%%%%%%%%%%%%%%%%%%%%%%%%%%

\paragraph{ Conclusions.---}

We have derived a determinant representation for the multivariate FCS of multicomponent impenetrable particles in the presence of a defect, 
following a quench from a bipartite setup. This framework provides access to the full transient dynamics via efficient numerical evaluation and 
enables the analytical computation of moments, cumulants, and multipoint correlation functions in the NESS in terms of scattering data and Jost 
solutions associated with the pre- and post-quench potentials.
Our results reveal qualitative features specific to multicomponent systems. In particular, low-temperature corrections are linear in temperature, 
in contrast to the quadratic behavior characteristic of single-component systems. We have also shown that bound states of the post-quench potential 
give rise to localized oscillatory contributions in densities and currents, which are absent when no bound states are present. 
For balanced systems at equilibrium, using the multicomponent generalization of the Levitov–Lesovik formula, we further find that the integrated 
spin current exhibits anomalous fluctuations. In the appropriate long-time scaling regime, its probability distribution is described by the 
M-Wright function.
Finally, the determinant representation in Eq.~\eqref{Lenard} extends, with minor modifications, to more general quench protocols 
involving arbitrary trapping potentials beyond localized defects, opening the way to a broader class of nonequilibrium applications.
Moreover,  since the determinant construction is not tied to a particular defect profile, it can be adapted to engineered barriers, quantum point contacts, and more general time-dependent trapping protocols \cite{PhysRevX.2.041020,PhysRevB.103.L041405}. It also supplies a microscopic benchmark for fluctuating hydrodynamics and for theories of anomalous spin transport \cite{Fujimoto2026,Derrida2007,PhysRevLett.131.197102,SciPostPhys.21.1.007}. Finally, the impenetrable limit provides a natural starting point for studying finite-interaction corrections, where spin exchange becomes dynamical and the interplay between coherent defect physics and interacting spin transport may generate new universality classes of nonequilibrium fluctuations.

%%%%%%%%%%%%%%%%%%%%%%%%%%%%%%%%%%%%%%%%%%%%%%%%%%%%%%%%%%%%%%%%%%%%%%%%%%%%%%%%%%%%%%%%%%%%%%%%%%%%%%%%%%%%%%%%%%%%%%%%%%%%%%%%%%%%%%%%%%%%%%%%
%%%%%%%%%%%%%%%%%%%%%%%%%%%%%%%%%%%%%%%%%%%%%%%%%%%%%%%%%%%%%%%%%%%%%%%%%%%%%%%%%%%%%%%%%%%%%%%%%%%%%%%%%%%%%%%%%%%%%%%%%%%%%%%%%%%%%%%%%%%%%%%%
%%%%%%%%%%%%%%%%%%%%%%%%%%%%%%%%%%%%%%%%%%%%%%%%%%%%%%%%%%%%%%%%%%%%%%%%%%%%%%%%%%%%%%%%%%%%%%%%%%%%%%%%%%%%%%%%%%%%%%%%%%%%%%%%%%%%%%%%%%%%%%%%

\begin{acknowledgments}
O.I.P. acknowledges financial support from Grant No.30N/2023, provided through the National Core Program of the Romanian Ministry of Research, 
Innovation, and Digitization. O.G. acknowledges support from Cognia AI.
C.P.M. acknowledges support by the Ministry of Education and Research,  CNCS/CCCDI-UEFISCDI, under project number 
PN-IV-P1-PCE-2023-0159 and PN-IV-P1-PCE-2023-0987 and by the Hungarian National Research, 
Development and Innovation Office—NKFIH Project No. K142179.

\end{acknowledgments}

\bibliography{FCS_V4.bib}

%%%%%%%%%%%%%%%%%%%%%%%%%%%%%%%%%%%%%%%%%%%%%%%%%%%%%%%%%%%%%%%%%%%%%%%%%%%%%%%%%%%%%%%%%%%%%%%%%%%%%%%%%%%%%%%%%%%%%%%%%%%%%%%%%%%%%%%%%%%%%%%%
%%%%%%%%%%%%%%%%%%%%%%%%%%%%%%%%%%%%%%%%%%%%%%%%%%%%%%%%%%%%%%%%%%%%%%%%%%%%%%%%%%%%%%%%%%%%%%%%%%%%%%%%%%%%%%%%%%%%%%%%%%%%%%%%%%%%%%%%%%%%%%%%
%%%%%%%%%%%%%%%%%%%%%%%%%%%%%%%%%%%%%%%%%%%%%%%%%%%%%%%%%%%%%%%%%%%%%%%%%%%%%%%%%%%%%%%%%%%%%%%%%%%%%%%%%%%%%%%%%%%%%%%%%%%%%%%%%%%%%%%%%%%%%%%%

 %                                                            END MATTER                                                                      %

%%%%%%%%%%%%%%%%%%%%%%%%%%%%%%%%%%%%%%%%%%%%%%%%%%%%%%%%%%%%%%%%%%%%%%%%%%%%%%%%%%%%%%%%%%%%%%%%%%%%%%%%%%%%%%%%%%%%%%%%%%%%%%%%%%%%%%%%%%%%%%%%
%%%%%%%%%%%%%%%%%%%%%%%%%%%%%%%%%%%%%%%%%%%%%%%%%%%%%%%%%%%%%%%%%%%%%%%%%%%%%%%%%%%%%%%%%%%%%%%%%%%%%%%%%%%%%%%%%%%%%%%%%%%%%%%%%%%%%%%%%%%%%%%%
%%%%%%%%%%%%%%%%%%%%%%%%%%%%%%%%%%%%%%%%%%%%%%%%%%%%%%%%%%%%%%%%%%%%%%%%%%%%%%%%%%%%%%%%%%%%%%%%%%%%%%%%%%%%%%%%%%%%%%%%%%%%%%%%%%%%%%%%%%%%%%%%
	
\clearpage
\setcounter{section}{0}
\setcounter{secnumdepth}{2}

\begin{center}
\textbf{End Matter}
\end{center}

%%%%%%%%%%%%%%%%%%%%%%%%%%%%%%%%%%%%%%%%%%%%%%%%%%%%%%%%%%%%%%%%%%%%%%%%%%%%%%%%%%%%%%%%%%%%%%%%%%%%%%%%%%%%%%%%%%%%%%%%%%%%%%%%%%%%%%%%%%%%%%%%
%%%%%%%%%%%%%%%%%%%%%%%%%%%%%%%%%%%%%%%%%%%%%%%%%%%%%%%%%%%%%%%%%%%%%%%%%%%%%%%%%%%%%%%%%%%%%%%%%%%%%%%%%%%%%%%%%%%%%%%%%%%%%%%%%%%%%%%%%%%%%%%%
%%%%%%%%%%%%%%%%%%%%%%%%%%%%%%%%%%%%%%%%%%%%%%%%%%%%%%%%%%%%%%%%%%%%%%%%%%%%%%%%%%%%%%%%%%%%%%%%%%%%%%%%%%%%%%%%%%%%%%%%%%%%%%%%%%%%%%%%%%%%%%%%

\section{Time-evolved orbitals}\label{timeevol}

In this section, we collect the necessary details for determining the exact form of the orbitals.
Let us recall the integral equation for the Jost function. For example, for the post-quench potential $V_F$, we obtain
\begin{align}\label{Jost}
\psi_k(z)
&=
e^{-i k z}-\int_z^\infty \frac{\sin[k(z-y)]}{\gamma k} V_F(y)\psi_k(y)\, dy, \\
\varphi_k(z)
&=
e^{-i k z}+\int_{-\infty}^z \frac{\sin[k(z-y)]}{\gamma k} V_F(y)\varphi_k(y)\, dy.
\end{align}
One can easily verify that these expressions satisfy the Schr\"odinger equation and have the correct asymptotics~\eqref{as1} and \eqref{as2}.
The corresponding Jost functions for the initial potential, $\Psi_k$ and $\Phi_k$, satisfy the same equations upon the replacement $V_F\to V_I$.
The initial orbitals are given by Eq.~\eqref{defL}. The time-evolved quantities can be found using the methods of Ref.~\cite{GZI23}.
The large-time asymptotics are presented in Eq.~\eqref{evLambda}. The coefficients $B^{(n)}_{S,q}$ are given by
\begin{equation}
B_{L,q}^{(n)}
=
\frac{i \Xi_{q,i\varkappa_n}^{L,\varphi} }{ a'_{i\varkappa_n}(q^2+\varkappa_n^2) }, 
\ \ 
B_{R,q}^{(n)}
=
\frac{i [\Xi_{q,i\varkappa_n}^{R,\psi}]^* }{  a'_{i\varkappa_n}(q^2+\varkappa_n^2)},
\end{equation}
where the coefficients $\Xi_{q,k}^{L,\varphi}$ and $\Xi_{q,k}^{R,\psi}$ are given by
\begin{align*}
\Xi_{q,k}^{L,\varphi}
&=
\Lambda_{L,q}'(0)\varphi_{k}(0) - \gamma\int_{-\infty}^0 \Lambda_{L,q}(z)V_D(z)\varphi_{k}(z)\, dz,
\\
\Xi_{q,k}^{R,\psi}
 &= 
 - \Lambda_{R,q}'(0)\psi_{k}(0) - \gamma\int_0^{\infty} \Lambda_{R,q}(z)V_D(z)\psi_{k}(z)\, dz.
\end{align*}
Here, $V_D(z)=V_I(z)-V_F(z)$. We can also write explicitly the $\mathcal{O}(t^{-1/2})$ contributions in Eq.~\eqref{evLambda}:
\begin{equation}\label{Se1}
\delta\Lambda_{S,q}(z,t)
=
\int_0^{\infty}\frac{dk}{\pi}\, C_{S,q,k}(z) \frac{e^{i t k^2/\gamma}}{(k+i0)^2-q^2},
\end{equation}
with 
\begin{align}\label{Sdefabc}
C_{L,q,k}(z)
&=\Phi_{q}(0)\mathrm{Re}\!\left[\frac{\Xi_{q,k}^{L,\varphi}\,\psi_k^*(z)}{\Phi_{q}(0)a_k}\right], \\
C_{R,q,k}(z)
&=\Psi^*_{q}(0)\mathrm{Re}\!\left[\frac{[\Xi_{q,k}^{R,\psi}]^*\varphi_k(z)}{\Psi^*_{q}(0)a_k}\right].
\end{align}
%

%%%%%%%%%%%%%%%%%%%%%%%%%%%%%%%%%%%%%%%%%%%%%%%%%%%%%%%%%%%%%%%%%%%%%%%%%%%%%%%%%%%%%%%%%%%%%%%%%%%%%%%%%%%%%%%%%%%%%%%%%%%%%%%%%%%%%%%%%%%%%%%%
%%%%%%%%%%%%%%%%%%%%%%%%%%%%%%%%%%%%%%%%%%%%%%%%%%%%%%%%%%%%%%%%%%%%%%%%%%%%%%%%%%%%%%%%%%%%%%%%%%%%%%%%%%%%%%%%%%%%%%%%%%%%%%%%%%%%%%%%%%%%%%%%
%%%%%%%%%%%%%%%%%%%%%%%%%%%%%%%%%%%%%%%%%%%%%%%%%%%%%%%%%%%%%%%%%%%%%%%%%%%%%%%%%%%%%%%%%%%%%%%%%%%%%%%%%%%%%%%%%%%%%%%%%%%%%%%%%%%%%%%%%%%%%%%%

\section{Currents and densities}\label{curandden}
	
The explicit expressions for the densities and currents follow immediately from the representation~\eqref{Lenard} and read
\begin{align}
\langle n_\sigma(x,t) \rangle 
&=
\sum_{S} 2\int_0^\infty \frac{dq}{\pi}\, \theta_S^\sigma(\epsilon_q)\, |\Lambda_{S,\epsilon_q}(x,t)|^2,\\ \nonumber
\langle j_\sigma(x,t) \rangle
&=
-\frac{4}{\gamma}\sum_{S} \int_0^\infty \frac{dq}{\pi}\,  \theta_S^\sigma(\epsilon_q) J_{S,q}(x,t), 
\\  J_{S,q}(x,t) 
& = \mathrm{Im}\left[ \partial_x\Lambda_{S,\epsilon_q}(x,t)\overline{\Lambda}_{S,\epsilon_q}(x,t) \right].
\end{align}
In the large-time limit, these expressions are given by Eqs.~\eqref{dens} and \eqref{current}, respectively.
The position-dependent coefficients $n_{\sigma,n}^{B}(x)$, $n_{\sigma,m,n}^{BB}(x)$, and $J_{\sigma,m,n}^{BB}(x)$ take the following form:
\begin{multline}
n_{\sigma,n}^{B}(x) = 2\int\limits_0^\infty \frac{ dq}{\pi} \left(
\theta^\sigma_L(\epsilon_q) |B_{L,q}^{(n)}\psi_{i\varkappa_n}(x)|^2 + \right. \\ \left.
\theta^\sigma_R(\epsilon_q) |B_{R,q}^{(n)}\varphi_{i\varkappa_n}(x)|^2 
\right).
\end{multline}
\begin{multline}
n_{\sigma,m,n}^{BB}(x) 
=
4\int\limits_0^\infty\frac{dq}{\pi} \theta^\sigma_L(\epsilon_q) [B^{(n)}_{L,q}]^*B_{L,q}^{(m)} \psi_{i\varkappa_n}(x) \psi_{i\varkappa_m}(x)  \\ 		
+
4\int\limits_0^\infty\frac{dq}{\pi} \theta^\sigma_R(\epsilon_q) [B^{(n)}_{R,q}]^*B_{R,q}^{(m)} \varphi_{i\varkappa_n}(x) \varphi_{i\varkappa_m}(x). 
\end{multline}
\begin{multline}
J_{\sigma,m,n}^{BB}(x)
= 
\frac{4}{\gamma}\int\limits_0^\infty\frac{dq}{\pi} \theta^\sigma_L(\epsilon_q) [B^{(n)}_{L,q}]^*B_{L,q}^{(m)} W[\psi_{i\varkappa_n}, \psi_{i\varkappa_m}](x)  \\ 		
+
\frac{4}{\gamma}\int\limits_0^\infty\frac{dq}{\pi} \theta^\sigma_R(\epsilon_q) [B^{(n)}_{R,q}]^*B_{R,q}^{(m)}W[ \varphi_{i\varkappa_n},\varphi_{i\varkappa_m}](x). 
\end{multline}
Here, $W[a,b](x)$ denotes the Wronskian of the two functions,
\begin{equation}
W[a,b](x) = \partial_x a(x) b(x) - a(x)\partial_x b(x),
\end{equation}
and we have used the fact that the bound-state wave functions are real and that the only complex part of $B^{(n)}_{S,q}$ is a phase independent
 of $n$.

%%%%%%%%%%%%%%%%%%%%%%%%%%%%%%%%%%%%%%%%%%%%%%%%%%%%%%%%%%%%%%%%%%%%%%%%%%%%%%%%%%%%%%%%%%%%%%%%%%%%%%%%%%%%%%%%%%%%%%%%%%%%%%%%%%%%%%%%%%%%%%%%
%%%%%%%%%%%%%%%%%%%%%%%%%%%%%%%%%%%%%%%%%%%%%%%%%%%%%%%%%%%%%%%%%%%%%%%%%%%%%%%%%%%%%%%%%%%%%%%%%%%%%%%%%%%%%%%%%%%%%%%%%%%%%%%%%%%%%%%%%%%%%%%%
%%%%%%%%%%%%%%%%%%%%%%%%%%%%%%%%%%%%%%%%%%%%%%%%%%%%%%%%%%%%%%%%%%%%%%%%%%%%%%%%%%%%%%%%%%%%%%%%%%%%%%%%%%%%%%%%%%%%%%%%%%%%%%%%%%%%%%%%%%%%%%%%

\section{Anomalous fluctuations of the spin current} \label{anofluctu}
	
In this section, we address the anomalous nature of the spin-current fluctuations associated with an arbitrary traceless Hermitian Cartan 
generator $\Lambda_z=\mathrm{diag}(q_1,\ldots,q_\kappa)$, with $\sum_{\alpha=1}^{\kappa}q_\alpha=0$, in thermal equilibrium. The corresponding 
spin (magnetization) in the right subsystem is $S_R^z(t)=\int_0^\infty\sum_{\alpha=1}^\kappa q_\alpha n_\alpha(x,t)\,dx$, and its characteristic 
function is
\[
\chi_{\subs}(\lambda;t)=\left\langle e^{i\lambda  S_R^z(t)} e^{-i\lambda  S_R^z(0)} \right\rangle .
\]
First, we consider the transferred charge. The corresponding FCS, $\chi_{\subc}(\lambda;t)\equiv \chi_R(i\lambda;t)$, is given by Eqs.~\eqref{LL00} 
and \eqref{LL0} (note that we use a complex counting field $i\lambda$). Its long-time asymptotics are given by the Levitov--Lesovik formula~\eqref{LL}, 
which we apply to the equilibrium case $\theta(\varepsilon)=\frac{\kappa}{\kappa+e^{\beta(\varepsilon-\mu)}}$. We arrive at 
\begin{equation}
\chi_{\subc}(\lambda,t) \simeq \exp\!\left[-\frac{\lambda^2}{2}\,\kappa_2^{\subc}(t)+ O(\lambda^4)\right],
\end{equation}
with $\kappa_2^{\subc}(t) = \hat{\kappa}_2^{\subc}\,t$ and  $\hat{\kappa}_2^{\subc}=\kappa/[\pi\beta(\kappa+e^{-\beta\mu})] $. 
Consequently, the distribution of the transferred charge is
\begin{equation}\label{pcm}
P_{\subc}(m,t) \simeq \frac{1}{\sqrt{2\pi\kappa_2^{\subc}(t)}} \exp\!\left[-\frac{m^2}{2\kappa_2^{\subc}(t)}\right].
\end{equation}

The key observation \cite{MPZD26} is that, for impenetrable particles, the spin sector remains completely disordered at all temperatures as the energy depends 
only on the charge degrees of freedom. Thus, each particle crossing the interface carries one of the $\kappa$ spin flavors with equal probability 
$1/\kappa$, irrespective of the direction of propagation. If $|m|$ particles cross the junction, the transported spin is the sum of $|m|$ independent 
spin weights. Each transferred particle carries a spin weight $q\in\{q_1,\ldots,q_\kappa\}$ with equal probability $1/\kappa$. The corresponding 
single-particle characteristic function is
\begin{equation}\label{ssc}
g_\kappa(\lambda)=\langle e^{i\lambda q}\rangle=\frac{1}{\kappa}\sum_{\alpha=1}^{\kappa}e^{i\lambda q_\alpha}.
\end{equation}
For a fixed number $m$ of transferred particles, the spin characteristic function is $[g_\kappa(\text{sign}(m)\lambda)]^{|m|}$. Averaging over the charge-transfer 
distribution $P_{\subc}(m,t)$ gives
\begin{equation}\label{sc}
\chi_{\subs}(\lambda,t)=
\sum_{m\in\mathbb Z} P_{\subc}(m,t) \left[g_\kappa(\text{sign}(m)\lambda)\right]^{|m|}.
\end{equation}
For small values of $\lambda$, we define
\begin{equation}\label{deflambda}
\phi_\kappa(\lambda) = -\ln g_\kappa(\lambda),
\ \ \phi_\kappa(\lambda)\underset{\lambda\to 0}{\sim}\frac{\lambda^2}{c_\kappa}.
\end{equation}
Here, $c_\kappa = 2/\sigma_{\subs}^2$, with $\sigma_{\subs}^2=\sum_{\alpha=1}^\kappa q_\alpha^2/\kappa$, is a positive constant determined by the choice 
of Cartan generator. The absence of a linear term in the expansion of  $\phi_\kappa(\lambda)$ follows directly from the tracelessness condition 
$\sum_{\alpha=1}^{\kappa}q_\alpha=0$.
	
In the long-time limit, using Eq.~\eqref{pcm} and replacing the sum in Eq.~\eqref{sc} by an integral, we obtain
\begin{align}
\chi_{\subs}(\lambda,t)
&\simeq
\int_0^{\infty}\frac{dm} {\sqrt{2\pi\kappa_2^{\subc}(t)}} \exp\!\left[-\frac{m^2}{2\kappa_2^{\subc}(t)}\right]
\nonumber\\
&\quad \times\left[e^{-m\phi_\kappa(\lambda)}+e^{-m\phi_\kappa(-\lambda)}\right].
\end{align}
For a general Cartan generator, $g_\kappa(-\lambda)$ need not equal $g_\kappa(\lambda)$. Tracelessness gives 
$\phi_\kappa(\pm\lambda)=\lambda^2/c_\kappa+O(\lambda^3)$, so both terms have the same leading contribution in the 
scaling limit. This expression reveals the non-Gaussian nature of the spin FCS: although the charge fluctuations are 
asymptotically Gaussian, the spin characteristic function depends nonanalytically on the number of transferred particles through $|m|$.
	
The probability distribution of the transferred spin is
\begin{multline}
\mathbb{P}_{\subs}[n,t]
=
\int\frac{d\lambda}{2\pi}\,e^{-i\lambda n}\,\chi_{\subs}(\lambda,t)= \\
\int\frac{d\lambda}{2\pi}\,e^{-i\lambda n}\int_0^\infty\frac{dm}{\sqrt{2\pi\kappa_2^{\subc}(t)}}
\exp\!\left[-\frac{m^2}{2\kappa_2^{\subc}(t)}\right] \\
\times\left[e^{-m\phi_\kappa(\lambda)}+e^{-m\phi_\kappa(-\lambda)}\right].
\end{multline}
Using the common leading term of $\phi_\kappa(\pm\lambda)$, introducing the scaling variables
\[
\lambda=\frac{\theta}{t^{1/4}},
\ \
m=t^{1/2}\mathcal{J}_c,
\ \
n=t^{1/4}\mathcal{J}_s,
\ \
\phi_\kappa(\lambda)
\sim
\frac{\lambda^2}{c_\kappa},
\]
we find
\begin{multline}
\mathbb{P}_{\subs}\!\left[t^{1/4}\mathcal{J}_s,t\right]
= \\
\frac{2}{t^{1/4}}\int_{-\infty}^{\infty}\frac{d\theta}{2\pi}\,e^{-i\theta\mathcal{J}_s}
\int_0^\infty\frac{d\mathcal{J}_c}{\sqrt{2\pi\hat{\kappa}_2^{\subc}}}
\exp\!\left[-\frac{\mathcal{J}_c^2}{2\hat{\kappa}_2^{\subc}}-\frac{\mathcal{J}_c\theta^2}{c_\kappa}\right].
\end{multline}
The Gaussian integral over $\theta$ can now be evaluated, leading to
\begin{multline}\label{wright}
\mathbb{P}_{\subs}\!\left[t^{1/4}\mathcal{J}_s,t\right]= \\
\frac{1}{t^{1/4}}\frac{1}{\pi}\left(\frac{c_\kappa}{2\hat{\kappa}_2^{\subc}}\right)^{1/2}
\int_0^\infty \frac{d\mathcal{J}_c}{\sqrt{\mathcal{J}_c}}\,
\exp\!\left[-\frac{\mathcal{J}_c^2}{2\hat{\kappa}_2^{\subc}}-\frac{\mathcal{J}_s^2}{2\mathcal{J}_c}\frac{c_\kappa}{2}\right].
\end{multline}
This proves Eqs.~\eqref{MW1} and \eqref{MW2}. Thus, the anomalous $t^{1/4}$ scaling of the integrated spin current follows from the 
combination of ordinary diffusive charge fluctuations, $m\sim t^{1/2}$, and the random spin composition of the transferred particles, 
which gives a typical spin fluctuation of order  $\sqrt{|m|}\sim t^{1/4}$. The resulting limiting distribution is the M-Wright distribution.

\end{document}